\documentclass[a4paper]{aa}
\usepackage{txfonts,graphicx,natbib}
\bibpunct{(}{)}{;}{a}{}{,}

\newcommand{\bz}{\ensuremath{\langle B_z\rangle}}
\newcommand{\bs}{\ensuremath{\langle B \rangle}}

\newcommand{\kms}{\ensuremath{\mathrm{km\,s}^{-1}}}
\newcommand{\vsi}{\ensuremath{v \sin i}}
\newcommand{\te}{\ensuremath{T_{\mathrm{eff}}}}
\begin{document}
\title{On the consistency of magnetic field measurements of Ap stars: lessons learned from the FORS1 archive}
       \author{
        J.D.~Landstreet  \inst{1,2}
       \and
        S.~Bagnulo      \inst{1}
       \and
       L.~Fossati	\inst{3}
       }
\institute{
           Armagh Observatory,
           College Hill,
           Armagh BT61 9DG,
           Northern Ireland, U.K.\\
           \email{jls@arm.ac.uk, sba@arm.ac.uk}
          \and
           Physics \& Astronomy Department,
           The University of Western Ontario,
           London, Ontario, Canada N6A 3K7. \\
%           \email{jlandstr@uwo.ca}
           \and
           Argelander Institut f\"{u}r Astronomie, 
           Auf dem H\"{u}gel 71, 
           Bonn D-53121, Germany
           \email{lfossati@astro.uni-bonn.de}
}

\authorrunning{J. D. Landstreet, S. Bagnulo, \& L. Fossati}
\titlerunning{The consistency of FORS1 magnetic field measurements}

\date{Received: 2014-03-21 / Accepted: 2015-03-22}
\abstract
% Context
{
The ESO archive of FORS1 spectropolarimetric observations may be used
to create a homogeneous database of magnetic field
measurements. However, no systematic comparison of FORS field
measurements to those obtained with other instruments has been
undertaken so far.
}
%Aims
{
We exploit the FORS archive of circular spectropolarimetric data to
examine in a general way how reliable and accurate field detections
obtained with FORS are. 
}
%Method
{
We examine the observations of Ap and Bp stars, on the grounds
that almost all of the unambiguous detections of magnetic fields in
the FORS1 archive are in these kinds of stars.  We assess the overall
quality of the FORS1 magnetic data by examining the consistency of
field detections with what is known from previous measurements
obtained with other instruments, and we look at patterns of internal
consistency.
}
%Results
{
FORS1 magnetic measurements are fully consistent with those
made with other instruments, and the internal consistency of the data is
excellent. However, it is important to recognise that each choice of
grism and wavelength window constitutes a distinct instrumental
measuring system, and that simultaneous field measurements in
different instrumental systems may produce field strength values that
differ up to 20\,\%, or more.  Furthermore, we found that field
measurements using hydrogen lines only yield results that 
meaningfully reflect the field strength as sampled specifically by lines 
of hydrogen for stars with effective temperatures above about 9000~K. 
%Finally, it is
%necessary to be aware of a problem of "occasional outliers" in the
%data.
}
%Conclusions
{
In general the magnetic field measurements of Ap and Bp stars obtained
with FORS1 are of excellent quality, accuracy and precision, and FORS1
provides an extremely useful example that offers valuable
lessons for field measurements with other low-resolution Cassegrain
spectropolarimeters.
}


\maketitle
%________________________________________________________________

\section{Introduction}
During a full decade of operations, between 1999 and 2009, the FORS1
instrument on the ESO Very Large Telescope collected a large number of
magnetic field measurements of various kinds of stars. Together with
the ESPaDOnS instrument of the Canada-France-Hawaii Telescope, and the
MuSiCoS and NARVAL instruments of the 2\,m Telescope Bernard Lyot of
the Pic-du-Midi Observatory, FORS1 has been one of the most important
instruments worldwide for observational studies of stellar magnetism.

Using the FORS1 instrument, magnetic fields have been detected in a
large number of previously unmeasured peculiar A and B (Ap and Bp)
stars \citep{Bagetal06,KocBag06,Hubetal06}, a class of stars long
known to harbour fields of typically kG strength . In addition, fields
have been securely discovered in a small number of other types of
stars such as the early B type $\beta$~Cep variable HD~46328 =
$\xi^1$~CMa \citep{Hubetal06}, the O6.5f?p star HD~148937
\citep{Hubetal08}, and several weak-field white dwarfs
\citep{Aznetal04,Lanetal12}. In contrast, a large number of rather
marginal magnetic field discoveries reported in non-Ap stars on the
basis of FORS1 measurements have been shown to be spurious
\citep{Bagetal12}. These are probably due to small instabilities in
the instrument, to image motion caused by atmospheric and seeing
variations, and to errors in the data reduction
\citep{Bagetal12,Bagetal13}. This situation makes clear the importance
of careful study of instrumental behaviour as an adjunct to scientific
measurements.

%Unfortunately, during the period when it was in use, very few
%calibration observations were obtained with FORS1 which could be used
%to securely establish the real level of uncertainties in
%measured polarisation values or of deduced mean longitudinal magnetic
%field strength \bz, and the degree to which magnetic field
%measurements made with FORS1 are consistent with one another and with
%field measurements with other spectropolarimeters.

%During the past decade, in a series of papers, serious efforts have
%been made to understand how best to treat magnetic measurements from
%FORS1. 
A first assessment of the capabilities of the instrument for measuring
fields in non-degenerate stars was made by \citet{Bagetal02}. This was
followed by development of methods to treat large numbers of
measurements in a consistent way \citep{Bagetal06}, and a careful
study of the general principles on which observing strategies should
be based \citep{Bagetal09}. Most recently, we have re-reduced the full
sample of circular spectropolarimetric data of stars with FORS1 in a
uniform way, and used the resulting large database to explore the
reproducibility and precision of those FORS field measurements used to
claim detection of weak magnetic fields \citep{Bagetal12}. Although
all of these studies have been focussed on FORS1, they are expected to
be generally relevant to field measurements with similar
low-resolution Cassegrain spectropolarimeters, and to apply in
particular to FORS2, which has inherited the polarimetric optics of
FORS1. 

However, no similar study has been undertaken of the quality and
behaviour of the magnetic field measurements in cases where the
detection of the field is not in question. This is the focus of the
present paper.

During the period when FORS1 was in use, over 1400 polarimetric
spectra were acquired of several hundred stars. Of the stars observed,
about 175 are known, at least provisionally, to have been classified
as magnetic Ap or Bp stars. More than 300 spectra of such stars were
obtained in polarimetric mode with FORS1. Such stars, if correctly
identified by classification, are known to possess magnetic fields
that can almost always be detected at the level of $\bz \sim 10^2$~G
or more \citep{Auretal07}. \bz, often called the mean longitudinal
field, is the mean value of the line-of-sight component of the
magnetic field, averaged over the visible stellar hemisphere, and is
the signature of a magnetic field to which a low-resolution
spectropolarimeter is sensitive. 

Thus it should be possible to study the characteristics of field
measurements with FORS1 in the strong-field limit (when detection of a
field is clear) by studying the internal consistency and systematic
behaviour of \bz\ measurements of single stars, and by comparing
observations of known magnetic Ap stars taken with FORS1 with similar
measurements obtained with other spectropolarimeters.

The following questions suggest themselves as ones that should be
examined: 
\begin{itemize}
\item Are measurements of \bz\ made with FORS consistent with
comparable measurements made with other spectropolarimeters?

\item FORS can be used with several different grisms and wavelength
regions. To what extent are \bz\ measurements obtained with these
different instrumental configurations similar or different?

\item It has become normal to report \bz\ values obtained with FORS
from spectral lines of hydrogen and from metal lines separately. To
what extent is this division meaningful, and what is the relationship
between the two kinds of measurement?

\item The basic approximation made in the deduction of \bz\ values
from FORS circular polarisation spectra is the so-called weak-field
approximation, which is expected to break down at large fields. Is
there evidence of such breakdown, either from comparison of \bz\
values derived separately from lines of hydrogen and of metals, or
from comparisons with field measurements with other instruments?
\end{itemize}

The next section of the paper describes the relevant features of
the data and the reductions. The third and fourth sections answer
the questions posed above as far as possible, and a final section
summarises our conclusions.

\section{Instrument settings and data reduction}

\subsection{Basic data reduction procedure}

The first experiments using FORS1 to measure fields in
non-degenerate stars \citep{Bagetal02} showed an obvious Zeeman
polarisation signal in the Balmer lines of the magnetic Ap star
HD~94660, and revealed that the forest of blended metal lines also
produces a complex signal in Stokes $V/I$. The corresponding value
of the mean longitudinal field \bz\ is obtained by using the
weak-field approximation (the approximation that the line splitting is
small compared to the width of the line): 
\begin{equation}
V/I = -g_{\rm eff} C_{\rm z} \lambda^2 \frac{1}{I} \frac{{\rm d}I}{{\rm d}\lambda}\bz,
\end{equation}
where $g_{\rm eff}$ is the effective Land\'e factor (which is 1.0 for
H lines and set to 1.25 for metal lines), $C_{\rm z} =
4.67\,10^{-13}$~\AA$^{-1}$, $I(\lambda)$ and $V(\lambda)$ are the
Stokes intensity and circular polarisation components from the stellar
polarisation spectrum, and $\lambda$ is wavelength \citep{Lan82}. This
equation is applied to each pixel in the $V(\lambda)/I(\lambda)$ spectrum, with
the required slope ${\rm d}I/{\rm d}\lambda$ estimated from the two adjacent
pixels. The resulting ensemble of values of $V/I$ are plotted as a
function of $\lambda^2 (1/I) ({\rm d}I/{\rm d}\lambda)$, a
straight line is fitted, and the mean value of \bz\ derived from the
slope of the correlation diagram. The uncertainty
in \bz\ is obtained from the uncertainty of the slope of the
correlation \citep{Bagetal12}.  

This process is illustrated in Figure~\ref{Fig_HD94660_reduction}. The
strong Zeeman signatures that coincide with the higher Balmer lines
are quite obvious in this figure, as is the fact that the $V/I$
spectrum is considerably "noisier" between the Balmer lines (in fact
with small Zeeman signatures from individual and blended metal lines)
than the expected photon noise, shown as a blue band below the
$V/I$ spectrum. The figure also shows for comparison a "null
spectrum", computed by co-adding individual polarisation sub-exposures
with signs changed in such a way as to cause the real polarisation
signal to cancel out. The null spectrum is a sort of "check spectrum"
that is expected to show about as much scatter as arises from photon
noise, and indeed this is what is seen \citep[e.g.][]{Bagetal12}.

This basic scheme has been used by virtually all observers to obtain
\bz\ values from FORS1 circular polarisation spectra. A number of
large surveys of various classes of stars have been published, several
of which have produced numerous marginal (3 to 5$\sigma$) detections
of magnetic fields \citep{Hubetal06,Hubetal09a,Hubetal09b},
particularly in classical Be stars and pulsating early-type B stars.
Combinations that have been used on FORS1 for magnetic measurements of
Ap stars are listed in Table~\ref{Tab_grism_choices}.

\begin{table}[ht]
\caption[]{\label{Tab_grism_choices}Grism -wavelength range choices
used on FORS1 to study Ap stars}
\begin{tabular}{rccccc}
\hline\hline
Grism   & Wavelength range & H lines \\
\hline
600\,R  &  4760 -- 6900 &  Ba-$\alpha$ to Ba-$\beta$ \\
        &  5250 -- 7400 &  Ba-$\alpha$               \\ 
600\,I  &  6900 -- 9050 &  Pa-10 to Pa-16            \\ 
600\,B  &  3470 -- 5880 &  Ba-$\beta$ to Ba-12       \\ 
1200\,B &  3670 -- 5130 &  Ba-$\beta$ to Ba-12       \\ 
1200\,g &  3840 -- 4970 &  Ba-$\beta$ to Ba-$\eta$   \\
        &  4290 -- 5470 &  Ba-$\beta$ to Ba-$\gamma$ \\
\hline\hline
\end{tabular}
\end{table}

\subsection{Error estimation}

Several of the weak field detection published by FORS users were in
conflict with clear non-detections with other instruments such as
ESPaDOnS at the CFHT \citep{Shuetal12}. This problem, together with
concerns about which of several possible reduction algorithms would
produce the most robust results, led us to build a mostly automated
reduction system around the ESO FORS pipeline \citep{Izzetal10}.
Thanks to this suite of software tools, almost the entire archive of
FORS1 magnetic measurements could easily be repeatedly re-reduced to
experiment with various algorithms and error estimates. This in turn
made it possible to identify the most nearly optimal methods of
reducing the FORS1 magnetic measurements and estimating realistic
measurement uncertainties.

We found that many surveys had underestimated the measurement
uncertainties of FORS1 data, and had not taken into account the
occurrence of small instrumental shifts due to flexures, and small
apparent radial velocity changes caused by seeing variations during
very short exposures \citep{Bagetal12,Bagetal13}. In general we found
that our error estimates were higher than those published in previous
works. We also identified the occurrence of "occasional outliers",
i.e., field measurements which appear significant at the 3 or
4$\sigma$ level but which are mainly due to small observational
velocity shifts rather than real Zeeman polarisation. Such outliers
also appear occasionally in the "null fields", field measurements
obtained from the null profiles, although the fields determined from
the null spectrum should be always found consistent with zero. The
occurrence of outliers led to the recommendation that field
discoveries with FORS should only be considered secure if they are
above about the $5\sigma$ level and are observed at least twice
\citep{Bagetal12}.

A major consequence of this project was to confirm (or at least not to
reject) a small number of marginal field detections from the surveys
in various classes of stars other than the magnetic Ap/Bp stars, but
to show that the large majority were spurious \citep{Bagetal12}.

A catalogue of \bz\ values for most of the FORS1 spectropolarimetric
observations is being made available by \citet{Bagetal14}.

%This large data reduction project has particularly focussed first on
%fully understanding the choices in principle in reducing data from
%instruments like FORS1 \citep{Bagetal09}, and secondly on optimal
%determination of the polarisation spectrum $V(\lambda)/I(\lambda)$,
%and on the accuracy with which the resulting uncertainty $\sigma_\bz$
%of the deduced \bz\ can be determined, in the context of deciding
%whether \bz\ is or is not significantly different from zero
%\citep{Bagetal12}.  Much less
%effort has so far gone into establishing the behaviour of \bz\
%measurements made with FORS1 for fields large enough that simple
%detection of a non-zero value is not an issue. We now turn to the
%analysis of such measurements, to the extent that they are available
%in the FORS1 spectropolarimetric archive, as a final step in this
%series of papers. 

\section{Characteristics of FORS1 \bz\ values derived from Ap stars}

\begin{figure*} \rotatebox{270}{\scalebox{0.600}{
\includegraphics*{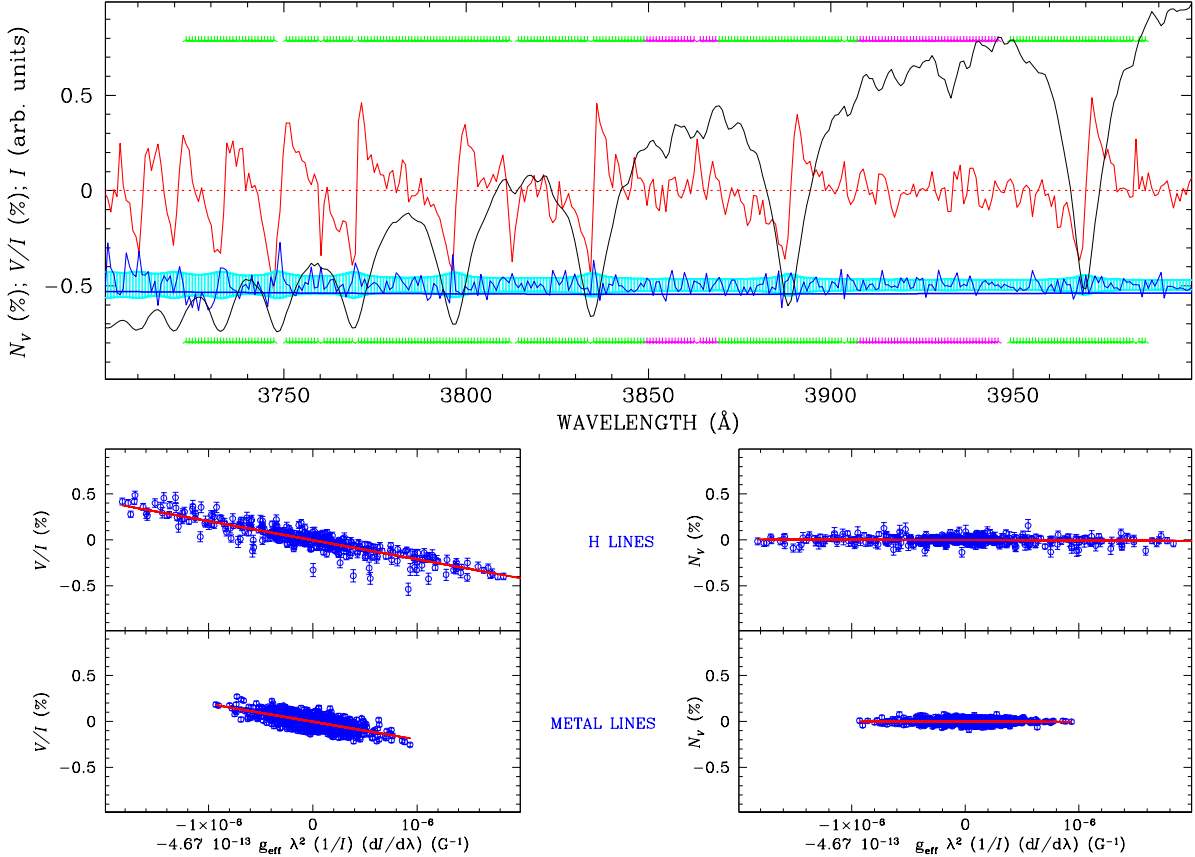}}}
\caption{\label{Fig_HD94660_reduction} Top panel: part of the intensity
spectrum of HD~94660, arbitrarily normalised (black); circular
polarisation spectrum $V(\lambda)/I(\lambda)$ (red); null
(verification) spectrum), shifted vertically by $-0.5$ for
clarity (blue); uncertainties of $V/I$ and null spectra, shifted
vertically by $-0.5$ for clarity (light blue band). The green stripes
above and below the spectrum show regions included in \bz\ measurement
in the H lines; pink stripes show regions included in the metal
line measurement. Lower left panels: correlation diagrams for
determination of field strength from $V/I$ and $I$ using H lines
(upper box) and using metal lines (lower box). Lower right panels: the
same with the null spectrum.  }
\end{figure*}

\begin{figure*}
\rotatebox{270}{\scalebox{0.600}{
\includegraphics*{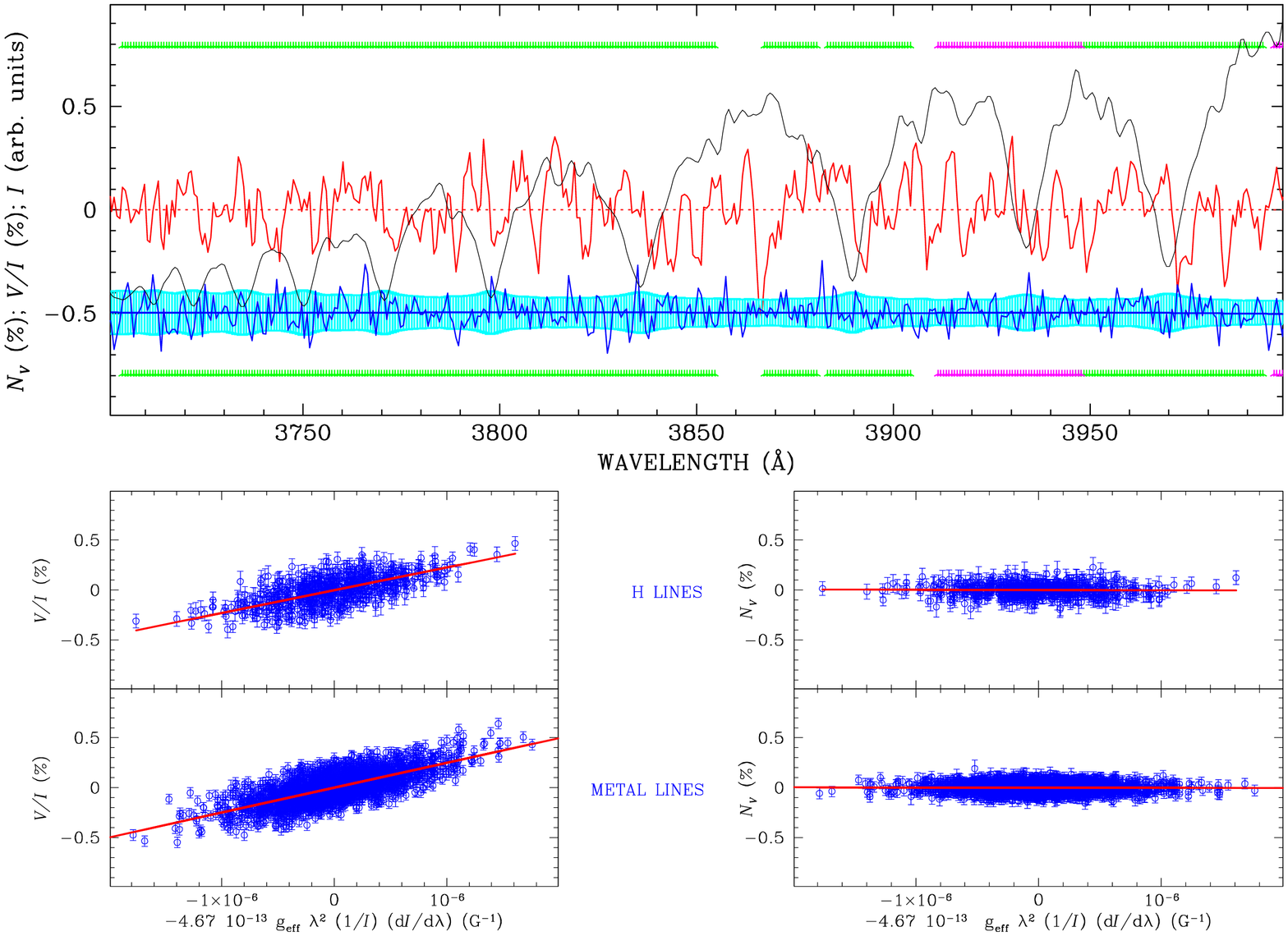}}}
\caption{\label{Fig_HD137949_no_H_signal} Portion of a spectrum in 
$I$, $V/I$, and
$N$ of the cool roAp star HD~137949. All symbols are the same as in
Figure~\ref{Fig_HD94660_reduction}. In contrast to that figure, notice
the absence of any systematic saw-tooth pattern in $V/I$ within the
Balmer lines.} 
\end{figure*}

This paper is devoted to understanding and characterising the
behaviour of FORS1 when to objective is to measure a magnetic field
whose existence is not in question. Because no observatory test or
calibration programmes were carried out specifically to address this
issue, we turn to the contents of the newly reduced FORS1 magnetic
field measurement archive to find data sets that allow us to study the
response of FORS1 to stellar spectra containing clear evidence of
Zeeman polarisation. As mentioned above, almost all the stars with
large enough fields to be easily detected belong to the class of
magnetic Ap and Bp stars. Therefore, from this point on, this paper
will almost entirely focus on FORS1 measurements of such stars.

\subsection{Measurements with various grism and wavelength combinations}

\begin{table}[ht]
\caption[]{\label{Tab_grism_comparison}Comparison of \bz\ values
obtained with different grisms}
\begin{tabular}{cccccr@{$\,\pm\,$}l}
\hline\hline
MJD         &  Grism      & $\lambda$ range &$R$ & $t_{\rm int}$ & \multicolumn{2}{c}{\bz (G)} \\
            &             & (\AA)           &    & (s)     & \multicolumn{2}{c}{(all lines)} \\
\hline \\
\multicolumn{7}{c}{HD 94660}\\
  54181.161 &    600\,B & $ 3470 -  5890$ & 1592 & 32      &$  -1952 $&    59 \\ 
  54181.147 &   1200\,B & $ 3800 -  4970$ & 2707 & 48      &$  -1998 $&    35 \\ 
\\
  52309.365 &    600\,B & $ 3480 -  5900$ & 1488 & 39      &$  -2567 $&    63 \\ 
  52309.375 &    600\,R & $ 5260 -  7420$ & 2175 & 73      &$  -2905 $&    56 \\ 
\\
  52383.122 &    600\,B & $ 3480 -  5900$ & 1616 & 80      &$  -2642 $&    59 \\ 
  52383.129 &    600\,R & $ 5250 -  7420$ & 2429 & 160     &$  -3224 $&    76 \\ 
\\
  53332.361 &    600\,B & $ 3470 -  5880$ & 1640 & 40      &$  -2552 $&    63 \\ 
  53332.374 &   1200\,g & $ 4290 -  5470$ & 2971 & 96      &$  -2576 $&    40 \\ 
\\
\multicolumn{7}{c}{HD 101065} \\
  52383.198 &    600\,B & $ 3480 -  5900$ & 1616 & 4780    &$  -1469 $&    97 \\ 
  52383.260 &    600\,R & $ 5250 -  7420$ & 2429 & 870     &$  -1445 $&    62 \\ 
\\
\multicolumn{7}{c}{HD 137949} \\
  52383.370 &    600\,B & $ 3480 -  5900$ & 1616 & 1289    &$   2689 $&    70 \\ 
  52383.408 &    600\,R & $ 5250 -  7420$ & 2429 & 320     &$   2843 $&    64 \\ 
\\
\multicolumn{7}{c}{HD 188041--42} \\
  52130.176 &    600\,B & $ 3480 -  5900$ & 1763 & 32      &$   2069 $&    66 \\
  52130.168 &    600\,R & $ 5350 -  7410$ & 2732 & 40      &$   2312 $&    61 \\
\\ 
\multicolumn{7}{c}{HD 201601} \\
  53976.260 &    600\,B & $ 3470 -  5880$ & 1725 & 60      &$  -1316 $&    64 \\ 
  53976.268 &   1200\,B & $ 3800 -  4960$ & 3052 & 122     &$  -1462 $&    31 \\ 
  52531.045 &    600\,R & $ 4760 -  6900$ & 2206 & 330     &$  -1658 $&    53 \\ 
  53335.011 &    600\,I & $ 6900 -  9050$ & 2956 & 16      &$   -596 $&    57 \\ 
\hline
\end{tabular}
\end{table}

The magnetic Ap and Bp stars are characterised by having fossil
fields. These are fields which change in structure only on time scales
much longer than $10^2$~yr, and which may therefore be considered to
have fixed surface magnetic geometries. The field structures are in
general not symmetric about the rotation axis, however, and so as a magnetic Ap or Bp star
rotates, the measured magnetic field, particularly the component along
the line of sight, varies periodically with the stellar rotation. 
 
Measurement of the value of \bz\ and its variation as a function of
the rotational phase of the underlying magnetic star generally leads
to a simple, usually sinusoidal, variation. The dispersion of the
individual data points with respect to the mean curve is determined by
the uncertainty of each measurement, and a strong test for the
correctness of the uncertainties is that the scatter about the mean
curve is consistent with them. Such {\it magnetic curves} are quite
useful for obtaining a simple, approximate model of the surface
magnetic field geometry of the star being observed. However, it is
well known that field measurements made with different instruments do
not lead to the same magnetic curves, or even the same extreme values
of \bz\ \citep[e.g.][]{BorLan77,Bycetal05}. Sometimes there appears to
be a modest scale change, or at times a shift of zero point.  

This effect is not really surprising. As discussed e.g. by
\citet{Lan82}, \bz\ is in principle a mean over the visible hemisphere
of local line of sight fields (or operationally of their polarisation
signatures), weighted by the local surface brightness and line
strength. Different elements will have different degrees of darkening
and line weakening towards the stellar limb, and this effect will
depend on the wavelength region used for observation. Thus the upper
Balmer lines will have a different weighting of the contribution of
various parts of the visible stellar hemisphere from disk centre to
limb than the upper Paschen lines, where the limb darkening is weaker
and the lines are shallower. Iron lines in the blue will have
different surface weighting than hydrogen, and than iron lines in the
red. Each specific choice of spectral lines to include in the
polarisation spectrum that is collapsed into a single \bz\ measurement
will lead to a somewhat different instrumental system in which to
measure \bz. Furthermore, many magnetic upper main sequence stars are
known to exhibit important variations in local abundances of various
chemical elements, both horizontally and vertically. This patchiness
varies (often quite strongly) from one element to another. Clearly,
the sampling of the magnetic field over the visible hemisphere of the
star will depend on the distribution of elements whose lines are used
for field measurement. This effect is expected to lead to differences
in the deduced value of \bz\ when this quantity is derived from
spectral lines of different elements, even within a single limited
spectral region. This effect has been observed: some Ap stars, such as
53~Cam = HD~65339 and $\alpha^2$~CVn = HD~112413, are known to yield
\bz\ measures that differ strongly depending on the element chosen for
the measurement, as shown by \citet[][Figures~4 and 5]{Wadetal00}.

The implications of this situation are clear: only by employing a
single instrumental observing system (grism, wavelength window,
resolving power), and by reducing all observations and deducing the
values of \bz\ with the same procedures, can we be fairly sure that
the magnetic curve(s) of a particular star will be simple and not
exhibit "excess" scatter. There is no guarantee that observations of a
single star that are obtained in different instrumental systems, or
reduced by a variety of methods, can be combined to produce magnetic
curves that will show the same degree of coherence, such as smooth
(even sinusoidal) variation with rotational phase. 
This is a very important consideration to keep in mind, because it is
easy to use FORS (and most similar instruments) in a variety of
instrumental configurations by changing the grism, the central
wavelength of observation, and the slit width. The ease with which a
variety of instrumental systems can be chosen has encouraged
observers to experiment with various settings, for example to see
which result in the most precise measurements of \bz\ per unit
time. It is clear from Table~\ref{Tab_grism_choices} that the
available choices may well have rather different weightings over the
stellar disk, and may also yield different values of \bz\ even for
simultaneous measurements. 

To explore how different are the resulting \bz\ values from these
different possible grism settings, we can look in the newly reduced
archive of measurements for stars with fairly large fields (1~kG or
more, to have good SNR for the field meaurements) and fields that
hardly change between measurements of \bz\ with different grisms,
either because the field measurements are back-to-back, or because the
period of field variation period is very long compared to the interval
between measurements. 

A list of the useful pairs found is shown in
Table~\ref{Tab_grism_comparison}, which provides the MJD of each
observation, the grism used, the range in wavelength of the spectrum
obtained, the resolving power $R$, the total shutter time $t_{\rm
int}$, and the \bz\ value measured. The field values in the table are
obtained using (almost) all lines in the spectral window. In each
case, the comparison is with the most commonly used configuration,
grism 600\,B. For the first four stars, the comparison measurements
were made immediately one after the other ("back-to-back"). For the
last set, taken with HD 201601 = $\gamma$~Equ, we use the fact that
the rotation period is of order 75~yr to make the approximation that
all four measurements refer to almost the same rotational phase.

We see from the comparisons that it is indeed important to check how
closely the same \bz\ values are obtained from simultaneous
measurements with different grisms if more than one instrumental
configuration is used. We also see that, in our test cases,
configurations with similar wavelength coverage as grism 600\,B
(grism 1200\,B and grism 1200\,g) yield very similar \bz\ values, but
that, compared to grism 600\,B, grism 600\,R seems to yield field values
about 10\% larger, while grism 600\,I (with only one case) appears to
provide a considerably smaller \bz\ value than is obtained with
grism 600\,B.

One other useful test we can make is to compare the final uncertainty
in \bz\ obtained in a given observing time using grism 600\,B and
either grism 1200\,B or 1200\,g. We have two back-to-back comparisons
between 600\,B and 1200\,B, one with HD~94660 and one with
HD~201601. In both cases for equal integration time the uncertainty
$\sigma_{\bz}$ with the higher dispersion grism is about 2/3 as large
as with the lower dispersion grism, suggesting that in this spectral
region there may be a substantial advantage to using the higher
dispersion grism. Even though grism 1200\,B covers a narrower
wavelength window than grism 600\,B, the amplitude of the $V/I$ signal
is larger, as is the slope ${\rm d}I/{\rm d}\lambda$, with the higher
dispersion grism, leading to more precisely determined values of \bz.

On the other hand, on the basis of a single comparison using HD~94660,
there appears to be no advantage to using grism 1200\,g (in the
wavelength window 4290--5470~\AA) rather than grism 600\,B (in its
usual wavelength window of 3470--5880~\AA). With equal integration
times the two grisms produce essentially equal uncertainties. This is
probably because the setting used for this observation with grism
1200\,g lacks all the higher Balmer lines, which in the usual setting
for grism 600\,B contribute quite a lot to the total signal, at least
for HD~94660.

\subsection{What is the meaning of "hydrogen" and "metal line" \bz\
measurements?}

The initial impetus to measure magnetic fields using lines of hydrogen
rather than those of common metals arose in part from the development
of hydrogen line photopolarimeters which were used in searches for
fields both in white dwarfs \citep{AngLan70} and in Ap stars
\citep{Lanetal75}. It was quickly realised that field measurements
made using hydrogen lines have two important advantages. First,
because hydrogen is overwhelmingly the dominant element in most
stellar atmospheres, a field measurement using H lines is unlikely to
be affected by significant non-uniformity of the abundance. That is,
hydrogen is the one element that probably samples the field
quasi-uniformly (except for limb effects) over the visible
hemisphere. Secondly, the Balmer lines are generally so broad that
sensitivity to a field is almost independent of \vsi, so that surveys
can be carried out that do not suffer from very strong selection
effects favouring low \vsi\ values.

The first experiments carried out on HD~94660 to explore magnetic
field detection in non-degenerate stars with FORS1 showed a clear
Zeeman polarisation signature in the Balmer lines, as expected.
Initially it was anticipated that the low resolution of FORS1 would
make field measurements using the heavily blended metal line spectrum
impractical, but experiments revealed useful polarisation signals in
the metal line spectrum between the Balmer lines
\citep{Bagetal02}. Ever since these first measurements, it has been
normal to report field estimates (made in the same way) separately for
the signal within the H lines, and for the (usually lower-amplitude
signal) in the continuum between Balmer lines. These are generally
called Balmer line and metal line measurements
respectively. Frequently a value of \bz\ is also reported using
(nearly) the full spectrum, including both Balmer lines and 
metal lines.

It is sometimes found that the Balmer line and metal line \bz\
measurements differ from one another, occasionally by a factor of up
to about two, while in many stars the two measurements are rather
similar. This situation has not been seriously discussed in the
literature, and it brings an air of uncertainty to measurements made
with FORS. It is certainly worth trying to understand the origin of
this effect.

We start by asking if the division during data reduction into H and
metal line measurements always has any
meaning. Figure~\ref{Fig_HD94660_reduction} shows an example of the
frequently used "standard star" HD~94660, in which a field of only
about $\bz \approx 2100$~G produces a strong, obvious saw-tooth Zeeman
pattern in the higher Balmer lines. In contrast, the amplitude of the
(significant) Zeeman polarisation in between the Balmer lines is
considerably lower, but of course can be exploited over a larger
wavelength window. In this star the division into H and metal line
measurements appears to be quite justified. In contrast, the spectrum
of the cool roAp star HD~137949 is shown in
Figure~\ref{Fig_HD137949_no_H_signal}. In this spectrum, a field of
almost exactly the same strength shows no clear Zeeman pattern
associated with Balmer lines at all. Instead, the signal in the $V/I$
spectrum, which is clearly much larger in amplitude than the the noise
spectrum seen in the $N_V$ spectrum, does not seem to be much
different within the Balmer lines than outside them. In this star, it
does not appear that there is a separately detectable Balmer line
signal without some very detailed modelling of the spectrum. In
HD~137949 the "Balmer line field" is actually simply the field as
measured using the metal lines that happen to be blended with H lines.

With this example in mind, it is worthwhile to try to find simple
criteria that enable us to know whether there is a meaningful \bz\
measurement that is primarily sensitive to the signal in H lines. In
stars with fields of a few kG or more, this can be determined by
direct inspection of the spectrum, as in the examples shown
here. However, for the far commoner case of a field of order 1~kG or
less, the polarisation signature is generally too weak in comparsion
to the noise to be classified visually. 

We have categorised a few stars with large fields as having strong
Balmer line signals or not on the basis of visual inspection. A
regularity that quickly emerges from this is that the stars with
effective temperatures above about 10000~K all have strong, obvious
Balmer line polarisation signals, while several roAp stars with \te\
around 7--8000~K and large fields all lack obvious signal in the H
lines. In the available FORS1 data, there are almost no strongly magnetic
Aps in the intermediate temperature range, but at a minimum we can
conclude that it is probably not specifically meaningful to assign
\bz\ value based on H lines to any roAp star, and probably not to any
Ap star with $\te < 9000$~K.

\subsection{Is there evidence of "saturation" of the metal line signal
for large \bz?}

\begin{figure}
\rotatebox{270}{\scalebox{0.350}{
\includegraphics*{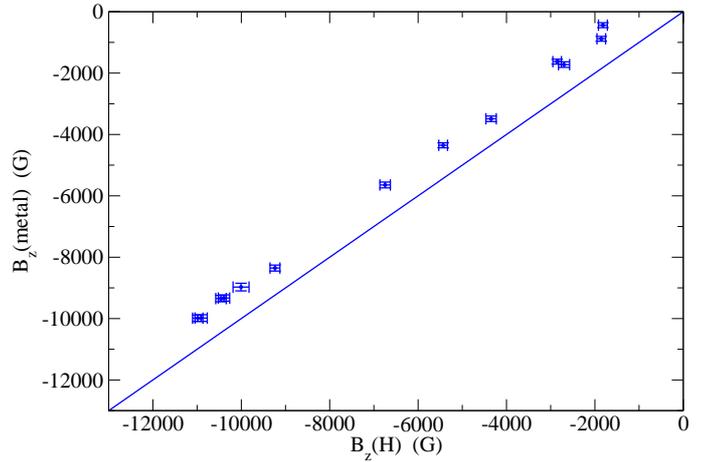}}}
\caption{\label{Fig_HD75049_H_vs_metal} Comparison of \bz\ measured
using Balmer lines (x-axis) with \bz\ measured using metal lines
(y-axis) in the strongly variable magnetic Ap star HD~75049. Each
point is the field measurement pair from a single FORS1 observation. } 
\end{figure}

\begin{figure} \rotatebox{270}{\scalebox{0.350}{
\includegraphics*{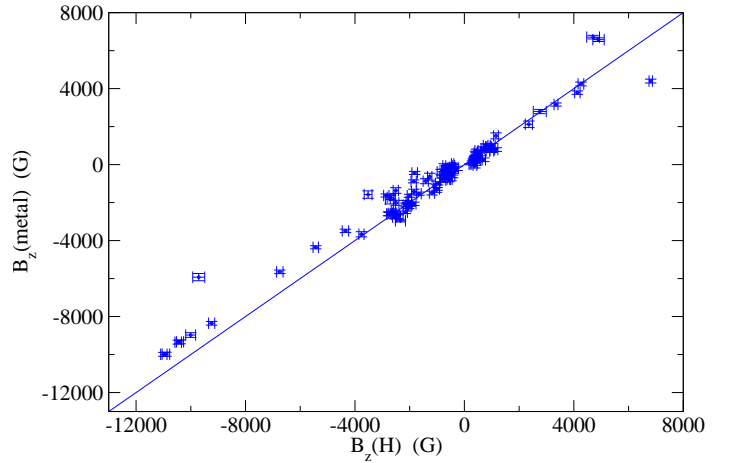}}}
\caption{\label{Fig_hot-aps-sig-flds-H-met} Comparison of \bz\
measured using Balmer lines (x-axis) with \bz\ measured using metal
lines (y-axis) in the full magnetic star sample, for stars with at
least $5\sigma$ field detections, with both H and metal $\sigma <
250$~G, and spectral types of early A or B. In the upper right corner,
the two outliers are HD~66318 with $\bz({\rm metal}) > \bz({\rm H})$, and
HD~318107 with $\bz({\rm metal}) < \bz({\rm H})$. In the negative field
region the two most obvious outliers are NGC 2244--334, well above the
mean line at $\bz({\rm H}) = -9700$~G, and NGC 2169--12 at $\bz({\rm
H}) = -3500$~G.  All remaining points with $\bz(H) < -4000$~G are
measurements of HD~75049. The line represents equality of the two \bz\
values. }
\end{figure}

One of the reasons that the division of \bz\ measurements into H line
measures and metal line measureswas made initially was the concern
that, because the determination of the value of \bz\ using metal lines
was based on the weak field approximation, the signal in metal lines
might cease to increase linearly with \bz\ at a field strength of only
1 or 2~kG: it might "saturate". In contrast, the polarisation in the
very broad H lines was expected to remain linear with \bz\ up to 10~kG
or more \citep{Bagetal06}. With a substantial sample of magnetic Ap
stars available, we can try to detect evidence of such saturation. 

One straight-forward method of searching for a breakdown in the
linear relationship between \bz\ as measured by H lines and \bz\ as
measured using metallic lines is to make use
of the fact that the hot star HD~75049 (with $\te \approx 10000$~K),
which has a very strong Balmer line Zeeman pattern visible in the
spectrum, has a field strength that varies from about 1 to 10~kG. We
can test whether the metal line \bz\ values begin to saturate in this
field strength interval by plotting \bz(H) versus \bz(metal) and
looking for non-linearity. Note that this test is possible because
this star has a high enough \te\ to have a clear H line Zeeman
signature; it could not be carried out with a star like HD~137949. 

The result of the comparison of \bz(H) vs \bz(metal) is shown in
Figure~\ref{Fig_HD75049_H_vs_metal}. It is clear from the figure that
there is a simple offset between the two kinds of \bz\ measurements (a
situation found from time to time in field measurements using various
other methods) but no non-linearity is apparent. It does not appear
that the values of \bz\ obtained from metal lines behave non-linearly
with respect to the values obtained from H lines up to at least
10000~G.  

Another useful test is to select hot stars from the full FORS1 Ap
sample that have clearly non-zero \bz\ values in both H lines and
metal lines, and compare the values. This result is seen in
Figure~\ref{Fig_hot-aps-sig-flds-H-met}. At first glance, this figure
shows a satisfactory level of agreement even up to $|\bz| \sim
10000$~G, but there is less new information, and more disturbing news,
than meets the eye. All the points with $\bz(H) < -4000$~G (except for
one strongly deviating point from NGC2244-334) are from one star,
HD~75049, already shown in Figure~\ref{Fig_HD75049_H_vs_metal}. A few
points lying close to (0,0) also show substantial difference between the
two values of \bz. Similarly, the two stars at
the far upper right of the figure deviate from the $\bz(H) = \bz$(metal)
line by a significant amount. However, overall there is no hint of
general departure from a linear relationship between \bz(metal) and the
field strength as measured by \bz(H).

Thus, our conclusion is that, within the limits of the available data,
there is no indication that the weak-field approximation used in
determining field strengths from FORS1 data saturates or becomes
non-linear, at least for field strength $\la 10$\,kG.

\subsection{Stars in which H line and metal line \bz\ values are
strongly discordant}

\begin{table*}[t]
\caption[]{\label{Tab_Espadons_Bz_values} \bz\ measures of 
ESPaDOnS spectra of stars with large \bz(H) -- \bz(metal) differences}
\begin{tabular}{cccccc}
\hline\hline
Star name     &   NGC\,2169-12 & \multicolumn{2}{c}{NGC\,2244-334} & HD\,318107  & HD\,49299  \\
ESPaDOnS file &   1656704      & 1046794         & 1671060         &  979157     &  1604088   \\
\hline
\bz(all)\ (G)  & $-2316 \pm 94$  & $-3940 \pm 172$ & $-4894 \pm 169$ &
   $4789 \pm 42$  & $-2371 \pm 18$ \\
\bz(Si)\ (G) & $-2781 \pm 216$ & $-3952 \pm 550$ & $-5560 \pm 487$ & 
   $7401 \pm 399$ & $-1275 \pm 96$ \\
\bz(Ti)\ (G)  & $-1769 \pm 507$ &                 &                 & 
   $4072 \pm 234$ & $-2833 \pm 136$ \\
\bz(Cr)\ (G)  & $-1678 \pm 226$ & $-3362 \pm 584$ & $-3840 \pm 541$ & 
   $4143 \pm 140$ & $-2497 \pm 23$ \\
\bz(Fe)\ (G)  & $-2688 \pm 104$ & $-4375 \pm 235$ & $-5398 \pm 201$ & 
   $5270 \pm 49 $ & $-2244 \pm 33$ \\
\hline\hline
\end{tabular}
\end{table*}

Although most of the measurements of \bz(H) and \bz(metal) agree
reasonably well (Figure~\ref{Fig_hot-aps-sig-flds-H-met}), a few hot
stars show quite large differences between these two kinds of
measurement. These unusual stars have differences of a factor of 1.5
or even 2 between the two values. Prominent examples of such
discrepancies are found in the FORS1 measurements of HD~66318,
HD~318107, NGC\,2244-334, and NGC\,2169-12. It is clear that these
differences may well be due to strongly non-uniform distributions of
some of the important elements, which then sample the field strength
differently over the surface than hydrogen does. 

We are able to test this hypothesis a little further because of the
availability of high-resolution polarisation spectra of three of
these discrepant stars taken with the ESPaDOnS spectropolarimeter on
the Canada-France-Hawaii Telescope. The ESPaDOnS spectra have
resolving power $R = 65\,000$, and cover almost the entire wavelength
window between 3800\,\AA\ and 1.04\,$\mu$m, and so it is possible to
measure the magnetic field using essentially only lines of one element
at a time (this is problematic for the FORS1 spectra because nearly
all metal lines are blended).

For some of the relevant ESPaDOnS spectra, field strengths of
useful precision can be derived from indiviual lines using by
measuring the separation of the line centroids in right and left
circular polarisation, or by the equivalent method of evaluating the
following expression across an isolated spectral line:
\begin{equation} 
\bz = 2.14 \times 10^{12} \frac{\int v V(v)dv}{\lambda g c \int[I_c -I(v)] dv}, 
\end{equation} 
where $v$ is a velocity coordinate across the line, $c$ is the speed
of light in the same velocity units, $\lambda$ is the unperturbed
wavelength of the line in \AA\ units, g is the Land\'{e} factor of the
line, and $I(v)$ and $V(v)$ are respectively the Stokes intensity and
circular polarisation components \citep{Mat89,Donetal97}. Note that
this method of evaluating \bz\ assumes that the spectral lines are
weak, but (unlike Eq.~(1)) it makes no assumption about the size of
the Zeeman splitting relative to the line widths, so it does not
saturate in the presence of large fields.

However, for most of the ESPaDOnS spectra, the signal-to-noise ratio
in $V$ is much too small for measurements precise enough to clearly reveal
differences in fields as sampled by the spectral lines of different
elements.  For this reason, values of \bz\ are derived for ESPaDOnS
polarised spectra by using the least-squares deconvolution method
\citep[LSD;][]{Donetal97}. The idea is to create averaged spectral
line and polarisation profiles, using the strong similarity of
different profiles (except for amplitude) when these are plotted in
velocity ($v$) space ($I(v)$ and $V(v)$). The contribution of each
line is weighted by the line strength, the wavelength and the
Land\'{e} factor, so that lines strongly sensitive to Zeeman splitting
are given most weight. This results in a single equivalent spectral
line intensity and polarisation profile which has much higher
signal-to-noise ratio than real single lines. The LSD line is assigned
an equivalent mean wavelength and Zeeman splitting factors, and it is
analysed as if it were a real single line, using Eq.~(2).

In the available ESPaDOnS spectra of stars from this discrepant sample
(NGC\,2169-12, NGC\,2244-334, and HD~318107), we are usually able to
measure \bz\ with useful accuracy using LSD average spectral
lines of the elements Si, Ti, Cr, and Fe. The results of this
experiment are shown in Table~\ref{Tab_Espadons_Bz_values}. (Note that
we cannot directly compare these field measurements to those obtained
with FORS1, as the ESPaDOnS spectra are in general from different
rotational phases than the available FORS1 spectra.) It is quite clear
from this table that in the stars showing a strong difference between
\bz(H) and \bz(metals) in the FORS data, there are also large and
clearly significant differences between the values of \bz\ measured
with lines of several individual elements. This certainly is
consistent with the idea that the discrepancy in the FORS measurements
is due to non-uniform distributions of various elements.

It is also interesting to carry out such an experiment for a star for
which the FORS \bz\ values measured with H lines and with metal lines
are in good agreement. HD~49299 ($\te = 9700$\,K, $\log g = 4.3$) is
such a star, with one measurement in which $\bz({\rm H}) = -2773 \pm
64$\,G while $\bz({\rm metal}) = -2589 \pm 104$\,G, and is a star for which
an ESPaDOnS spectrum is available. The results of measuring \bz\ with
lines of various elements are also shown in
Table~\ref{Tab_Espadons_Bz_values}, where it is seen that even in this
star the value of \bz\ measured using lines of Si is strongly
different from the values obtained using lines of iron peak elements!
It appears that patchy distributions of various elements are probably
very common in magnetic Ap stars, and that accurate agreement between
\bz(H) and \bz(metal) usually depends on the structure of the patches
rather than on their absence.

\section{Comparisons of measurements of fields of well-studied magnetic Ap stars}

The FORS1 archive includes some field measurements of Ap stars
that have been observed with other instruments, or that have been
observed often enough that internal consistency tests can be carried
out.

It is well known that the absolute values of measurements of the same
stars obtained with different instruments display noticeable
discrepancies. Examples of this effect may be found in \citet[][ see
particularly Figs.~14, 17 and 38]{Mat91}, and in \citet[][particularly
Figs.~3, 4, 5, 9 and 10]{Wadetal00}, where the extremely high precision
of the MuSiCoS observations make the various discrepancies quite
clear.

As discussed in Sect.~2.2, such disagreements are due to several
causes. The basic problem is that the way that the circular
polarisation signal from which \bz\ is deduced is sampled over the
visible stellar hemisphere depends on the wavelength window and
resolving power used, on the varying effects of limb darkening and
line weakening on lines of different wavelengths, excitation
potentials and strengths, and on possible horizontal and vertical
abundance variations of the element(s) whose lines are included in the
averaged polarisation signal.

Nevertheless, the \bz\ values as measured with FORS1 should be similar
in strength (usually within perhaps 20\,\%) to \bz\ values measured at
the same phase in the field variation cycle with other
methods. Furthermore, the variations of \bz\ observed with different
methods should generally be similar in amplitude and phase. Thus the
comparisons we will make are useful to establish that no major
difference exists between the data coming from FORS1 and from other
instruments. The new reductions of the FORS1 data can also be used in
several cases discussed below to study the internal consistency and
repeatability of FORS1 magnetic measurements.

\subsection{The ``reference star'' HD\,94660}

\begin{figure} \rotatebox{0}{\scalebox{0.350}{
\includegraphics*{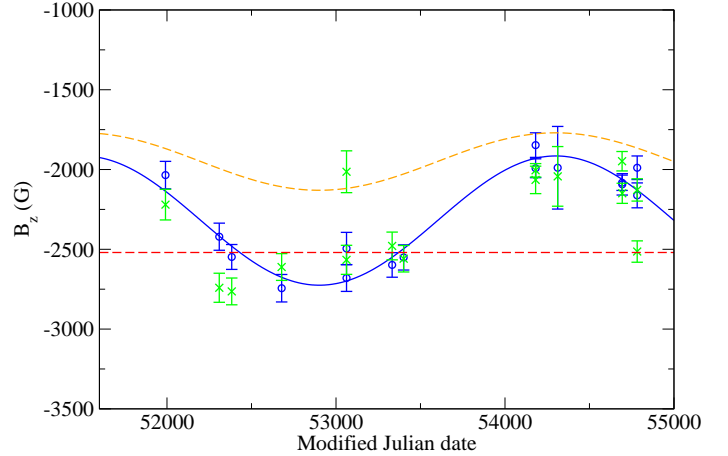}}}
\caption{\label{Fig_HD94660-vs-MJD} Variation of \bz\ for HD~94660
with MJD as measured using Balmer lines (blue error bars with
circles), and metallic spectrum (green error bars with x's), for data
obtained using the 600\,B and 1200\,B grisms. The solid smooth curve
is best sine wave fit to \bz\ measured using Balmer lines, assuming $P
= 2800$\,d. The dashed straight red line is the approximately
constant field reported by \citet{Bohetal93}; the dashed orange curve
is the sine wave fit to CASPEC \bz\ measurements described by
\citet{LanMat00}.}  \end{figure}

HD~94660 = HR~4263 is a southern magnetic Bp star of $V =
6.11$. Its effective temperature \te, derived from Geneva and
Str\"{o}mgren photometry \citep{Meretal97} with the {\tt FORTRAN} codes
developed by \citet{Napetal93} and \citet{Kunetal97} (as described in
more detail by \citet{Lanetal07}) is about 11500~K, and $\log g
\approx 4.1$.

The field of HD~94660 was one of the first fields discovered using the
Balmer-line magnetograph \citep{BorLan75,Bohetal93}. The rotation
period is approximately 2800~d, determined from the small variations
of the mean field modulus between 6.05 and 6.4\,kG
\citep{LanMat00,MatHub06}. The longitudinal field, as measured by
\citet{MatHub97} using the CASPEC instrument at ESO La Silla Observatory with
polarimetric optics, varies by less than about 400~G full amplitude,
between about $-1750$ and $-2100$\,G, but at present the details of
this variations are still not very clear because of the small
amplitude of variation relative to the measurement uncertainties
\citep{LanMat00}.  Nevertheless, the small variation in \bz\, together
with a declination of $-42\degr$, makes this star a convenient
reference for checking polarimetric optics in the southern hemisphere.

HD\,94660 was observed with FORS1 17 times between 2002 and 2008, with
grisms 600\,B, 600\,R, 1200\,g, and 1200\,B. The star was occasionally
observed quasi-simultaneously with more than one grism, and these
field measurements help us to understand how the \bz\ field diagnostic
changes using different regions of the optical spectrum, as discussed
in Sec. 3.1 above. HD~94660 has strong Zeeman signatures in $V/I$ in
each of the Balmer lines, so it is of interest to examine how well the
\bz\ values obtained using H lines agree with \bz\ values from metal
lines. From the 14 measurements using the two common grism
instrumental systems, 600\,B and 1200\,B, 13 show reasonable agreement
between H line and metal line \bz\ values, and do not show any
significant trend for one value to be larger than the other. On one
night (MJD=54782.376, grism 1200\,B) the \bz\ values in the two
systems differ by slightly more than $3 \sigma$. This may be an
example of the ``occasional outlier'' problem that we have identified
in FORS1 data, as discussed in Sect. 2.2 above. On two other nights
differences of a little less than $3\sigma$ are observed.

A second useful test is the consistency of repeated observations made
during one night with a single grism choice. We have one such double
observation using grism 600\,B, and two using grism 1200\,B. In these
three cases, all pairs of Balmer line measurements are completely
consistent, as are two of the three metal line measurements. However, the
pair of metal line measurements with grism 1200\,B from MJD=54782.376
differs by almost $4 \sigma$, reflecting the same outlier mentioned
above.

We next consider the relationships between \bz\ values measured in
various instrumental systems.  The field of HD~94660 was measured
simultaneously on one night using both the instrumental system of
grism 600\,B and that of grism 1200\,B (see
Table~\ref{Tab_grism_comparison} above). There is no significant
difference between the \bz\ values measured in the two systems, which
suggests that -- for this star at least -- we may be able to combine
the data from these two measurement systems to study the field
strength variations. \bz\ measurements in grism 600\,B and grism
1200\,g on a single night also agree within the uncertainties, so --
again for this star -- these two systems appear to be approximately
comparable.

In the two measures using grism 600\,R, the Balmer line measures are
about 25\,\% lower, and the metal line measures about 25\% larger, than
field values measured on the same night with grism 600\,B.  For HD\,94660,
these data can clearly not be combined with those from other
instrumental systems without determining the substantial adjustment
required.

When we fit a sine wave to the Balmer line \bz\ values obtained with
the 600\,B and 1200\,B grisms, we find a clear periodicity with a
period of $2800 \pm 250$\,d, and no other acceptable period in the
range of 1000 to 5000~d. The variation of \bz\ with time as measured
separately with Balmer lines and metal lines, in the grism 600\,B and
1200\,B systems, is shown in Figure~\ref{Fig_HD94660-vs-MJD}, together
with the fit to the Balmer line data. The clear detection of the
period found independently in the \bs\ data \citep{MatHub06} is strong
confirmation that the observed variations are stellar rather than
instrumental. However, both the mean \bz\ value (about $-2300$\,G) and
the amplitude of variations (about 800~G peak-to-peak) are somewhat
larger than the range of \bz\ values shown by \citet{LanMat00}, so
that combining the FORS1 data with the earlier CASPEC measurements
would require rescaling of one of the two data sets.

The best-fit sine wave to the CASPEC \bz\ data discussed by
\citet{LanMat00} is shown in Figure~\ref{Fig_HD94660-vs-MJD}. The
CASPEC data are based on polarised spectra typically recorded in the
wavelength window 5400--6800\,\AA, considerably to the red of the
window used for most of the FORS1 data, and so we expect some offset
in the mean values of \bz, as seen in the Figure. Nevertheless, after
modest shifting and re-scaling, the agreement is satisfactory. Note
that this star has very sharp spectral lines, so observations with a
high-resolution spectropolarimeter, such as ESO's HARPSpol, should be
able to provide an extremely high-precision magnetic curve -- in yet
another instrumental system.

Our conclusion is that observations of the standard Ap star HD~94660
are consistent with previous data, and are quite similar to the
approximately constant value of $\bz \approx -2520$\,G found by
\citet{Bohetal93} using the Balmer-line magnetograph (note sign error
in their Table~1), as indicated in
Figure~\ref{Fig_HD94660-vs-MJD}. However, there is a substantial scale
difference between the FORS1 \bz\ measurements and those obtained with
CASPEC. The variations observed in the FORS1 \bz\ data set are
consistent with the known period of variation of the star, and are
presumably real. This star remains a very useful standard for testing
polarimetric instruments and systems in the southern hemisphere.

\subsection{The very strongly magnetic star HD 75049}

\begin{figure}
\rotatebox{270}{\scalebox{0.350}{
\includegraphics*{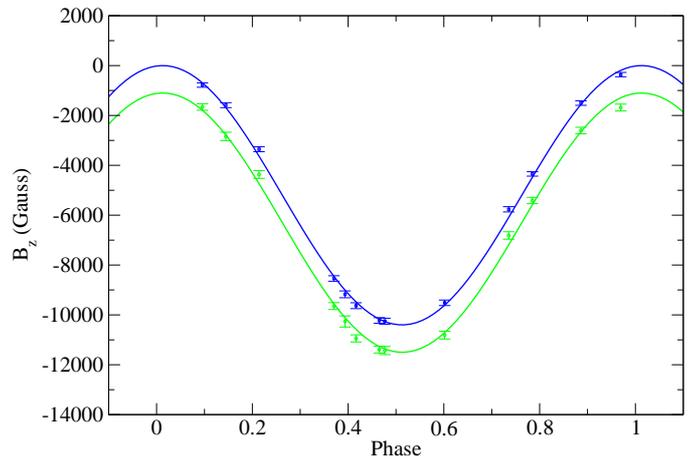}}}
\caption{\label{Fig_hd75049-phase} Variation of \bz\ with rotational
  phase for HD~75049 as measured using Balmer lines (lower data
  set), and metallic spectrum (upper data
  set), for all FORS1 \bz\ data. The smooth curves
  are sine wave fits to the two \bz\ curves, assuming
  $P = 4.0476$~d.  }
\end{figure}
 
HD~75049 = CPD--50~1770 is a relatively unknown ninth magnitude
southern Ap star. According to \citet{Elketal10} it has $\te = 9600$~K
and $\log g = 4.1$. Its distance is unknown.

A magnetic field with $\bs \sim 30$~kG was discovered in the star
HD~75049 by \citet{Freetal08}. The star has been studied in detail by
\citet{Elketal10}, who obtained 13 \bz\ magnetic field measurements
using FORS1 with the 600\,B grism during a period of about 3
months. Using this material, 
\citet{Elketal10} concluded that the rotation period of the star is
$P = 4.04899 \pm 0.00008$\,d, and that both \bz\ and \bs\ vary smoothly
and approximately sinusoidally with that period. Although \bz\ does
not change sign, the amplitude of variation is quite large, about
10\,kG between extrema.

From our data reduction we estimate that a typical standard error of
the field as measured with metal lines is larger by roughly a factor of
two relative to the measurements reported by \citet{Elketal10}. Our field
measurements are also slightly different from those reported by
\citet{Elketal10}, but not by important amounts. This star is hot
enough that the polarisation signal due to hydrogen Balmer lines is
quite clear, so that measuring the field in Balmer lines has a
physical meaning. Accordingly we discuss the metal and Balmer line
measurements independently.

We have first re-determined the best period using the re-reduced FORS1
magnetic measurements. Compared to the data used by \citet{Elketal10},
the larger error bars in our data lead to lower values of the reduced
chi-square value ($\chi^2/\nu$) of the best-fit sine wave, around 2.5
to 3.0, compared to a value of about 8 for the \bz\ values found by
\citet{Elketal10} for the spectral range 3212--6215\,\AA. However, our
best estimate of the rotation frequency and uncertainty from the
re-reduced metal line data is $f = 0.2471 \pm 0.0002$~cycles~d$^{-1}$,
or a period of $P = 4.047\pm 0.003$~d, consistent with but about 30
times less precise than the value reported by \citet{Elketal10}.

We have also examined the \bs\ data reported by \citet{Elketal10} for
Fe~{\sc ii} 5018~\AA\ to try to improve the period we derive. This is
the line for which the variations of \bs\ have the largest amplitude
and apparently have the least scatter. When these data are fit with a
sinusoid, the resulting period is about $P = 4.0475 \pm 0.0058$, even
less accurate than the period derived from the FORS1 \bz\ data.

We conclude that the period of this star can only be determined from
the published magnetic data to a precision of about $P = 4.047 \pm
0.003$~d.

From the relatively small value of $\chi^2/\nu$ that we find for the
fit of a simple sine wave to the FORS1 \bz\ measurements, we
conclude that our standard errors are not underestimated by very much;
in fact, if they are accurate, the small remaining discrepancy between
the sine wave and the observations could be due to a very slightly
non-sinusoidal variation of \bz. 

Both the metal line \bz\ data and the H line \bz\ values can be fit
about equally well with sine waves, with compatible best periods, but
the \bz\ values found using H lines are systematically about 1000\,G
more negative than those found using metal lines, as was already found
by \citet{Elketal10}. This difference is shown in different ways in
Figures~\ref{Fig_HD75049_H_vs_metal} and
Figure~\ref{Fig_hd75049-phase}, where the variation of \bz\ as
measured in the two different instrumental systems is illustrated. The
fact that the metal line data are well fit by a simple sine wave, and
that the sine wave fits to \bz\ measures using metal lines and using H
line are essentially the same except for a zero-point shift, give
further evidence that the metal line \bz\ measurements vary
approximately linearly with field strength in this field strength
range, up to at least about 10~kG.  As discussed by \citet{Elketal10}
and above, the systematic difference between the \bz\ values measured
using the metal lines and those measured with H lines, which appears
to correspond to a vertical shift of about 1100\,G, is probably due to
the fact that the H lines sample the visible disk differently than the
typical metal lines. This systematic difference is typical of
measurements made of \bz\ using different instrumental systems.

\subsection{The roAp star HD~83368}
The star HD~83368 = HR~3831 = HIP~47145 is (barely) bright enough
to be in the Bright Star Catalogue. It is one of the first Ap stars
discovered to show the short-period pulsations characteristic of the
rapidly-oscillating Ap (roAp) stars \citep{Kur82}. Using photometry to
estimate its basic properties as we did for HD~94660 above, we find
$\te \approx 7500$~K and $\log g \approx 4.4$. This star is cool
enough that, although the Zeeman polarisation signature is quite clear
in the star, there is no systematic difference between the appearance
of this signal in and out of the Balmer lines. We do not think that
there is any particular physical meaning that can be attached to \bz\
values measured in the Balmer lines, and so we discuss only the \bz\
values determined from the full spectrum.

HD~83368 was observed twice on one night, first with grism 600\,B over
a prolonged period during which more than 120 frames were collected at
a rate of roughly one per minute, and later with grism 600\,R for a
few minutes \citep{Hubetal04}. Our archive includes only the average
results of each of these two data sets.

We notice that the two \bz\ values from the full spectrum, $\bz =
+1018 \pm 30$~G (600\,B) and $\bz = +893 \pm 52$\,G (600\,R) are
consistent with one another. The small difference may be due to
statistical noise, or to the fact that the two grism setting sample
different wavelength windows (3480 -- 5900\,\AA\ and 5250 -- 7420\,\AA\
respectively), and thus sample the field strength over the disk
somewhat differently.

To make a comparison with previous field measurements, we use the
ephemeris of \citet{MatHub97}. Assuming that the last digit of the
period is significant, we find that the measurement with grism 600\,B
was taken at $\phi = 0.52 \pm 0.01$, which corresponds to the positive
extremum of the field. According to the magnetic curve of
\citet{Tho83}, the field measured with a Balmer line filter
polarimeter would be about $+725$\,G at this phase. According to the
fit to CASPEC spectropolarimetry, the field at this phase should be
about $+560$\,G. Thus the FORS1 \bz\ measurement is significantly
larger than would be found with either of these other two methods, but
the difference is not much larger than the difference present between
the two comparison values, and is within the range of variation found
between different field value measurement methods.

\subsection{The extremely peculiar roAp star HD 101065}

HD~101065 = HIP 56709 is Przybilski's star, the most peculiar of the
peculiar A stars known. Its spectrum is dominated by lines of such
rare earth elements as Dy and Ho. It is the coolest known magnetic Ap
star, with $\te = 6400$\,K and $\log g = 4.2$ \citep{Shuetal10}. The
star has very sharp lines ($\vsi = 3.5 \pm 0.5$\,\kms \citet{Cowetal00}), but variations are
small and the rotation period is unknown. It has a magnetic field
which was discovered by \citet{WolHag76} from three photographic
Zeeman measurements. All three measurements were in the range between
$-2000$ and $-2500$\,G. Recent observations by \citet{Hubetal04} using
FORS1 have confirmed the presence of the field, but their measurements
(long series of short exposures taken on two different nights to
search for field variations with pulsation) yielded \bz\ field
measurements of about $-1000$\,G. Hubrig et al. remark that the field
measured in the Balmer lines is about 500\,G larger than in the metal
lines, a feature we confirm. However, as discussed above, this star is
so cool that the "Balmer line \bz\ values" are simply
measurements of the field using metal lines which are sampled
differently over the spectrum from the usual metal line measurement.

There are altogether 13 observations of HD~101065 in the FORS1
archive. Twelve have been taken with grism 600\,B in very similar
wavelength windows, so these data form a set taken in an essentially
fixed instrumental system. Thus these 12 measurements are directly
comparable to one another. The twelve 600\,B observations cover about
five years with a variety of intervals between measurements, so it is
of interest to see if the \bz\ data reveal the rotation period of the
star. A periodogram was computed for the metal line data for all
distinguishable periods between 10~yr and 1 day (assuming longer
periods are ruled out by the non-zero \vsi\ value reported by
\citet{Cowetal00}). The fit spectrum shows that the spread of the
\bz\ values is not large enough to reliably find a period: the values
of $\chi^2/\nu$ for a sinusoidal fit to the data range only between
0.4 and 2.9. The best periods appear to be around 4.30 and 1.79\,d,
but it is not at all clear that one of these is the actual rotation
period, assuming that the rotation period is actually within the
range tested.

In fact the dispersion in the \bz\ data relative to the overall mean,
about 105\,G, is only marginally larger than typical standard error of
about 70\,G, and the reduced chi square value of fit assuming that the
field is constant is about 1.35.  Thus, these data do not reveal
strong evidence that the field of HD~101065 is detectably
variable. Possibly the star is observed nearly pole-on, consistent
with the small value of \vsi, or perhaps the field axis is
aligned with the rotation axis. Alternatively, it is possible
that the reported value of \vsi\ is actually indistinguishable from
zero; the star is a roAp pulsator, and the very small reported \vsi\
is certainly close to the threshold for reliable detection in a
pulsating, strongly magnetic star. In this case, HD~101065 might have
an extremely long rotation period of decades, like HD~201601, and the
decade of field measurements with FORS1 may simply cover only a small
fraction of the rotation period.

The mean \bz\ value measured with the Balmer line windows is about
$-1850$\,G, while that measured in the usual way with with metal lines
is about $-1250$\,G. The Balmer line \bz\ values correspond more
closely to the values observed by \citet{WolHag76} than do the metal
line values. In spite of the fact that this star is so cool that
H does not make a strong contribution to the spectropolarimetric
signal within Balmer lines, this difference between the two
measurement values is one of the larger differences found among Ap
stars. It probably reflects considerable non-uniformity in the metal
lines that dominate the field measurement, or the peculiarity of the
atmospheric structure of HD~101065, and illustrates clearly the
difficulty in interpreting simple field moment measurements such as
\bz.

\subsection{The very long period roAp star HD 137949 = 33 Lib}

\begin{figure} \rotatebox{0}{\scalebox{0.350}{
\includegraphics*{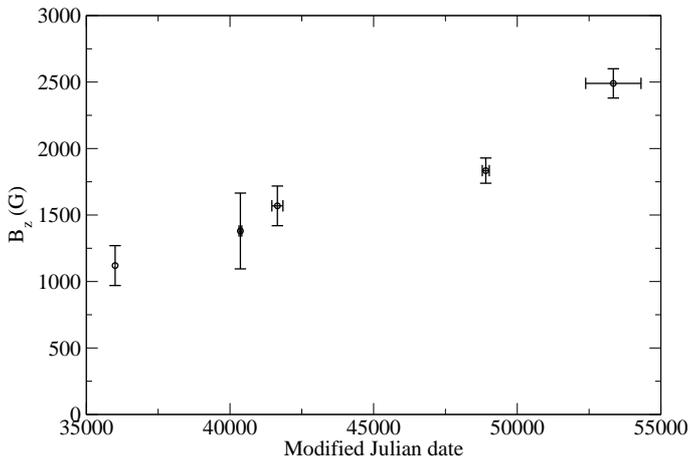}}}
\caption{\label{Fig_Bz_long_term_hd137949}Variation of mean \bz\
values for HD~137949 as a function of MJD between 1957 and 2007. Each
point represents the mean field measured by a single observer or team,
with the vertical error bar showing the dispersion of individual
measures (or an estimate of the uncertainty for the first point), and
the horizontal error bar showing the extent in time of the measurement
set from that team. }
\end{figure}

HD~137949 = HIP~75848 = 33~Lib is a cool SrEuCr magnetic Ap with
$\te = 7400$~K, $\log g = 4$ and a low mass, probably about $1.7
M_\odot$ \citep{Netetal08,Shuetal13}. It is also a roAp pulsator
\citep{Kur82}. It has very sharp spectral lines with clearly resolved
Zeeman splitting visible in Fe~{\sc ii} 6149~\AA. The mean
longitudinal field \bz\ has been measured by a number of investigators
since the field was discovered by \citet{Bab58}, as summarised by
\citet{Matetal97}. The value of \bz\ appears to have been increasing
more or less steadily over the past half century at a rate of the
order of 20~G/yr. The mean field modulus \bs\ was measured
repeatedly during the 1990s over 4.3~yr \citep{Matetal97}, and was
found to remain absolutely constant (within uncertainties of roughly
$\pm 40$~G) at 4660~G.

These facts are consistent with a very long rotation period, of the
order of $10^2$~yr, and the secular change of \bz\ as measured
over the decades (see Figure~\ref{Fig_Bz_long_term_hd137949})
certainly suggests such a long period. However, it is important to
keep in mind that each of the points in the Figure was obtained with a
different instrumental system. The first three points, representing
the mean value of \bz\ observed by respectively \citet{Bab58},
\citet{vandenHeu71} and \citet{Wol75}, were obtained photographically
with three different telescope--polarisation analyser--coud\'{e}
spectrograph combinations (the Palomar 5-m, the Lick 3-m, and the
Mauna Kea 2.2-m). All these observations were made in the blue,
typically with Kodak 103a-O plates covering roughly 3800--4900~\AA,
and so might seem to be on essentially a single instrumental
system. Comparison of the Lick and Mauna Kea observations of standard
stars suggest that both systems are closely comparable and yield
similar \bz\ data \citep{WolBon72}, but it appears that measurements
made at Lick or Mauna Kea are roughly 20\,\% larger than Mount
Wilson--Palomar measurements \citep{PrePyp65}. Thus within
uncertainties the first three \bz\ data points in the Figure may well not
be significantly different.

The \bz\ measures contributing to the fourth point, at about MJD
48900, were obtained in a different wavelength window, between about
5600 and 6800\,\AA\ \citep{Matetal97}. From our exploration of FORS1
data, we can anticipate that the different mix of elements
contributing important spectral lines, and the different limb
darkening, could easily lead this point to be shifted by 20\,\% or so
from the scale of the first three points. Thus the reality of the
secular trend apparent in the first four data groups is not securely
established.

The field of HD~137949 was measured seven times over about 5~yr with
FORS1, in all but one case with essentially the same instrumental
setup using grism 600\,B. Five of the six 600\,B measurements were
carried out within a period about three months in 2007, while the
first measurement with 600\,B was made five years earlier, in
2002. (One \bz\ measurement with grism 600\,R was obtained in 2002 as
well, yielding a \bz\ value only a little larger than the simultaneous
measurement with grism 600\,B; see Table~\ref{Tab_grism_comparison}.) 
If we measure the standard deviation of the full spectrum measurements
using all six 600\,B values, the dispersion found is about two times
larger than would be predicted for a constant field and accurate
uncertainties. The dispersion drops to about the expected value if we
omit the 2002 measurement. The dispersion of the five values measured
in 2007 supports our estimates of the standard errors; the 2002 FORS1
\bz\ value using all lines is roughly $3\sigma$ larger than the mean
of the other five values.

The simplest interpretation of this result is that the field has
actually decreased in strength by roughly 230~G in the five years
separating the 2002 FORS1 measurement from the others. Of course, if
the old \bz\ values are all in compatible measurement systems, so that
the earlier measurements really do show an increasing field during the
second half of the 20th century, this field may have now reached
maximum and be declining again. Alternatively, it is possible that the
first \bz\ measurement with the 600\,B grating is one of the
occasional data outliers.

To place the FORS1 data in the context of the long-term changes
displayed in Figure~\ref{Fig_Bz_long_term_hd137949}, note that the
average FORS1 data point was obtained in a wavelength window roughly
centred on, but a little wider than, the wavelength windows used for
the three points from photographic measurements. We would not expect a
large offset of the FORS1 \bz\ values relative to those obtained using
the photographic instrumental systems. It appears that the FORS data
support the reality of the very slow secular change in \bz\ over half
a century at a rate of the order of 30~G~yr$^{-1}$, and consequently
an extremely long rotation period, of many decades.

\subsection{The very strongly magnetic, periodically variable roAp star
  HD 154708}

\begin{figure}
\rotatebox{0}{\scalebox{0.350}{
\includegraphics*{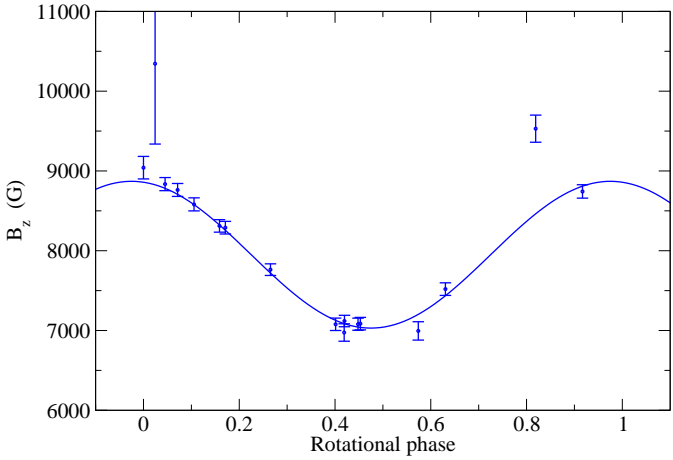}}}
\caption{\label{Fig_hd154708-phase} Variation of \bz\ with rotational
  phase for HD~154708 as measured using the full spectrum and data
from both grism 600\,B and 1200\,g.  The smooth curve
  is the best sine wave fit to \bz, assuming
  $P = 5.363$~d.  }
\end{figure}

A large field was discovered in the very cool Ap star HD~154708, which
has $\te = 6800$~K, $\log g = 4.1$, and $M = 1.5 M_\odot$, by
\citet{Hubetal05}. There are 17 \bz\ measurements of the star in the
FORS1 archive. Of these, 14 were taken within an interval of 110~d,
all with the 600\,B grism and covering the same wavelength interval,
so that these 14 measurements are in a single instrumental system.  In
addition, three earlier measurements are available. Two were taken
with the 1200\,g grism, covering about half the wavelength interval of
the 600\,B measurements with about twice the spectral resolution, and
one was taken with 600\,B in a wavelength window slightly different
from that used in the long data series, so that it is in essentially
the same instrumental system as the 14 contiguous measurements. All
these field measurements have been published, with slightly different
\bz\ values than in our catalogue for each point, by
\citet{Hubetal09c}. They report a rotation period of $P = 5.3666 \pm
0.0007$~d. Note that this star is so cool that H line measurements
have no particular meaning, so we discuss \bz\ data obtained using all
lines.

It is found from the series of measurements taken with the 600\,B grism
that the value of \bz\ varies by a total of about 25\% of the mean
value of \bz\ during rotation. This is small enough that, as discussed
above, the relationship of the two \bz\ measurements taken with the
1200\,g grism relative to the larger sample of data is uncertain by an
important fraction of the total. These data cannot be used to
refine the period of variation without considering the relationship of
the two different instrumental systems. 

We have two pieces of information that suggest that it may be possible
to safely combine the data from the 1200\,g grism with that from the
600\,B. First, the one comparison that we have of this grism with the
600\,B, two almost simultaneous measurements of \bz\ of HD~94660 on
MJD 53332.3 (see Table~\ref{Tab_grism_comparison}), yields essentially
identical \bz\ values in the two systems. Unfortunately, the
wavelength setting used for the 1200\,g grism is considerably to the
red of the setting used for grism 600\,B, so this result may not apply
to all Ap stars, especially if their temperatures are very different
from HD~94660.

The second piece of circumstantial evidence that the two sets of data
can be combined is the fact that the values of the two measurements
made with the 1200\,g grism by chance coincide with the largest and
smallest values of \bz\ observed with grism 600\,B. This strongly
suggests that \bz\ values measured with the two grisms cover a very
similar range of values, and thus {\em are} reasonably comparable. 

The most recent 13 FORS1 measurements, using all lines, define
clearly the rotation period of this star. A periodogram of these 13
data points yields a unique period of $5.369 \pm 0.013$~d with a
minimum $\chi^2/\nu$ value of about 0.56. 

We can try to refine this period using the two remaining measures
obtained with grism 600\,B. Adding these points makes the minimum value
of $\chi^2/\nu$ rise to about 3.2, and two minima suggest two equally
possible periods, $5.386 \pm 0.004$ and $5.362 \pm 0.004$.  These two
best periods are near the two extremes of the coarser period defined
by the 13 point sample. Thus we cannot improve significantly on the
period defined by the 13 most recent contiguous measurements.

We then introduce the two earlier field measurements made with grism
1200\,g, and carry out a sine wave fit on all 17 data points. The best
value of $\chi^2/\nu$ rises to 3.25, and the best period is between
5.3596 and 5.3671, or $P = 3.3634 \pm 0.0038$~d. Most of the increase
in the value of $\chi^2/\nu$ comes from the first (grism 600\,B) point
in 2007, so we drop this point and fit the remaining 16 points,
including the two measurements made with grism 1200\,g. The unique
best fit, with values of $\chi^2/\nu$ as low as 0.93, is found for
periods in the range of 5.3600 to 5.3665. This corresponds to $P =
5.363 \pm 0.003$~d, almost exactly the value found with all 17 points.

We note that the period of $5.3666 \pm 0.0007$~d proposed by
\citet{Hubetal09c} on the basis of all the FORS1 \bz\ measurements is
consistent with the one derived here. However, we are unable to
reproduce from these data the very small uncertainty assigned by
\citet{Hubetal09c}. We believe that they have underestimated the
uncertainty of the rotation period by about a factor of 4.

The variation as a function of rotational phase of the 15 grism 600\,B
and the two grism 1200\,g \bz\ values, determined using all lines, is
shown in Figure~\ref{Fig_hd154708-phase}. The smooth curve is a sine
wave fit to the 16 "good" \bz\ values obtained using all lines. In
general, the data points fit the best fit curve very well. This
overall good fit suggests that our uncertainties
are realistic. The exception to the good fit (the deviating data
points at $\phi = 0.84$) illustrates yet again the occasional outliers
found in FORS data.

\subsection{The long-period roAp star HD 210601 = $\gamma$ Equ}
The \bz\ field component of this star has been measured frequently
over about 60 years. The data are summarised (and supplemented) by
\citet{Bycetal06}, who show that (with a dispersion of roughly 500~G,
probably due to the variety of measurement methods adopted) the \bz\
data are consistent with a period of 80 or 90~yr. This long period has
been confirmed (although its value is still quite uncertain) by the
slow, regular change of position angle of linear polarisation of the
star \citep{Leretal94}. \bz\ appears to have been near negative
extremum during the past decade, with a value of about $-1000$~G.

The FORS1 catalogue includes four different measurements of this star,
each one with a different grism, different resolving power, and
different wavelength region. The star is sufficiently cool that the H
measurement is not meaningful, so we discuss only the \bz\ values
obtained from all lines. These data are shown in the last entries of
Table~\ref{Tab_grism_comparison}. The measurements scatter over the
same range found at this rotational phase with other methods, from
about $-450$ to about $-1700$~G. The typical value is about $-1200$~G,
also similar to the results of other measurements.

Let us assume that in fact the magnetic field structure on the visible
hemisphere of the star changed little during the four year span in
which the observations were made, so that the formally defined value
of \bz\ has hardly changed during that period. What
Table~\ref{Tab_grism_comparison} really illustrates is the point we
have made repeatedly above, namely that the FORS spectropolarimeter
(like similar low-resolution spectropolarimeters) is easily used in a
variety of different instrumental modes, each with its own sensitivity
to chemical abundance, limb darkening, line weakening, and stellar
parameters, and each providing a different numerical value of \bz\
even when the mathematical definition of this quantity from the data
is unchanged.  In this star particularly we see that when searching
for periodicity or studying the time variation of \bz, it is essential
to use only a single instrumental system.

\section{Conclusions}

1. The overall \bz\ scale used for FORS1 is realistic, similar to
(although generally not identical to) the \bz\ scales resulting from
other field measurement methods.\\

\noindent 
2. The standard errors of measurement as estimated by our new
reductions are realistic. Fields are generally detected in stars where
they are already known to exist, or are repeatedly and/or convincingly
detected. The dispersion of \bz\ values around mean variation curves
are generally consistent with the standard errors we find, apart from
the problem of "occasional outliers". \\

\noindent
3. The division of \bz\ measurements into hydrogen line and metal line
measurements is quite meaningful for Ap stars hotter than perhaps
9000\,K. Below this temperature, the distinctive Zeeman signatures in
the Balmer lines are overwhelmed by metal line polarisation, and
hydrogen \bz\ values are not meaningfully different from the metal
line measurements except for the fact that they sample the overall
wavelength window differently. For cool Ap stars, the hydrogen line
measurements have no strong connection with the field as sampled
by hydrogen. \\

\noindent 4. The field strengths measured with FORS1 are no more
different from previous measurements than typical differences between
various instrumental measurement systems already in wide
use. Furthermore, the values of \bz\ as measured using Balmer lines,
when these are meaningful, are usually reasonably close to those
measured using the metallic spectrum. Differences between the hydrogen
line and metal line instrumental systems are usually not large even
when \bz\ approaches 10\,kG, or \bs\ reaches 30\,kG. Since these
field strengths are close to the upper limits observed in any magnetic
Ap or Bp stars, it appears that there is no problem with using the
weak  field approximation of Eq.~(1), which is expected to
break down at "very large" fields, to derive \bz\ from the metallic
spectrum of any magnetic main sequence star observed with FORS1. 
However, we do find a few ``occasional outliers'' that should agree
with others but instead deviate by 3 or even $4\sigma$ from the
expected values. \\

\noindent 
5. Each specific grism and wavelength setting used with FORS defines a
specific instrumental system for measuring \bz. Because of difference
in sensitivity to patchiness, and differences in limb darkening and
line weakening, measurements made with these different systems are
{\em not} directly comparable except for general magnitude. In
particular, it will generally not be safe to use measurements in one
instrumental system to refine a period derived with another
instrumental system, because of possible (usually unknown) \bz\ value
shifts or scale changes between the systems, until the relevant shifts
have been determined with adequate precision.\\

\noindent 
6. These conclusions are all expected to apply, more or less exactly,
to other similar low-resolution, Cassegrain-mounted
spectropolarimeters, and specifically to FORS2, but also, for example,
to the ISIS spectropolarimeter of the William Herschel Telescope. \\

\noindent
7. FORS in polarimetric mode provides a valuable and extremely
powerful tool for measuring magnetic fields in stars, provided its
characteristics are well understood. In particular, it is important
not to underestimate errors, to be aware of the "occasional outlier"
problem, to not expect to reach standard errors of 20 or 30\,G without
a lot of extra care, and to carry out programmes studying time
variation with a single instrumental system.

\acknowledgements{ We thank the referee, Dr G. Mathys, for his careful
reading and very useful report. The data used in this work have been
provided by the ESO Science Archive Facility from the following
programmes: 060.A-9203, 068.D-0403, 069.D-0210, 070.D-0352,
072.C-0447, 073.D-0464, 074.C-0442, 074.D-0488, 075.D-0295,
077.D-0556, 079.D-0240, 079.D-0697, 079.D-5023, 080.D-0170, 081.D0670,
082.D-0342.  The data used were obtained in part from
observations carried out at the Canada-France-Hawaii Telescope (CFHT),
which is operated by the National Research Council of Canada, the
Institut National des Science de l'Universe of the Centre National de
la Recherche Scientifique of France and the University of Hawaii.
Work on this project by JDL has been supported by the Natural Sciences
and Engineering Research Council of Canada. LF acknowledges support
from the Alexander von Humboldt Foundation.}


\begin{thebibliography}{}

\bibitem[Angel \& Landstreet(1970)]{AngLan70} Angel, J. R. P.,
Landstreet, J. D. 1970, ApJ, 160, L147

%%%\bibitem[Appenzeller(1967)]{App67} Appenzeller, I. 
%%%                          1967, PASP, 79, 136

%%%\bibitem[Appenzeller et al.(1998)]{Appetal98} Appenzeller, I., Fricke, K., Furtig, W., et al. 
%%%                          1998, The Messenger, 94, 1

\bibitem[Auri\`ere et al.(2007)]{Auretal07} Auri\`{e}re, M, Wade,
G. A., Silvester, J., et al. 2007, A\&A, 475, 1053

\bibitem[Aznar Cuadrado et al.(2004)]{Aznetal04} Aznar Cuadrado, R., Jordan, S., Napiwotzki, R., 
                          Schmid, H.M., Solanki, S.K., \& Mathys, G.
                          2004, A\&A, 423, 1081

\bibitem[Babcock(1958)]{Bab58} Babcock, H. W. 1958, ApJS, 3, 141

\bibitem[Bagnulo et al.(2002)]{Bagetal02}
                          Bagnulo, S., Szeifert, T., Wade, G.A., Landstreet, J.D., \& Mathys, G. 
                          2002, A\&A, 389, 191

\bibitem[Bagnulo et al.(2006)]{Bagetal06} Bagnulo, S., Landstreet, J. D., Mason, E., et al.\
                          2006, A\&A, 450, 777

\bibitem[Bagnulo et al.(2009)]{Bagetal09} Bagnulo, S., Landolfi, M., Landstreet, J.D.,
                          Landi Degl'Innocenti, E., Fossati, L., \& Sterzik, M. 
                          2009, PASP, 121, 993

\bibitem[Bagnulo et al.(2012)]{Bagetal12} Bagnulo, S., Landstreet,
  J. D., Fossati, L., Kochukhov, O. 2012, A\&A, 538, A129

\bibitem[Bagnulo et al.(2013)]{Bagetal13}Bagnulo, S., Fossati, L.,
Kochukhov, O., Landstreet, J. D. 2013, A\&A 559, A103

\bibitem[Bagnulo et al.(2014)]{Bagetal14}Bagnulo, S, Fossati, L.,
Landstreet, J. D. 2014, in preparation. 

\bibitem[Bohlender et al.(1993)]{Bohetal93} Bohlender, D. A.,
  Landstreet, J. D., Thompson, I. B. 1993, A\&A, 269, 355

\bibitem[Borra \& Landstreet(1975)]{BorLan75} Borra, E. F.,
Landstreet, J. D. 1975, PASP, 87, 961

\bibitem[Borra \& Landstreet(1977)]{BorLan77} Borra, E. F.,
Landstreet, J. D. 1977 ApJ 212, 141

\bibitem[Bychkov et al.(2005)]{Bycetal05} Bychkov, V. D., Bychkova,
L. V., Madej, J. 2005, A\&A 430, 1143

\bibitem[Bychkov et al.(2006)]{Bycetal06} Bychkov, V. D., Bychkova, L. D., \& Madej, J. 2006, MNRAS, 365, 585

\bibitem[Cowley et al.(2000)]{Cowetal00} Cowley, C. R., Ryabchikova, T., Kupka, F. et al. 2000, MNRAS, 317, 299

\bibitem[Donati et al.(1997)]{Donetal97} Donati, J.-F., Semel, M., Carter, B. D., Rees, D. E., 		Cameron, A. C. 1997, MNRAS, 291, 658

\bibitem[Elkin et al.(2010)]{Elketal10} Elkin, V. G., Mathys, G.,
Kurtz, D. W., Hubrig, S., Freyhammer, L. M. 2010, MNRAS, 402, 1883

\bibitem[Freyhammer et al.(2008)]{Freetal08} Freyhammer, L. M., Elkin,
V. G., Kurtz, D. W., Mathys, G., Martinez, P. 2008, MNRAS, 389, 441

\bibitem[Hubrig et al.(2004)]{Hubetal04} Hubrig, S., Kurtz, D.W., Bagnulo, S., Szeifert, T., Sch\"{o}ller, M.,
                          Mathys, G., \& Dziembowski, W.A. 
                          2004, A\&A, 415, 661

\bibitem[Hubrig et al.(2005)]{Hubetal05} Hubrig, S., Nesvacil, N.,
  Schoeller, M. et al. 2005, A\&A 440, 37

\bibitem[Hubrig et al.(2006)]{Hubetal06} Hubrig, S., Briquet, M., Sch\"{o}ller, M., De Cat, P.,
                          Mathys, G., \& Aerts, C.
                          2006, MNRAS, 369, 61

\bibitem[Hubrig et al.(2008)]{Hubetal08} Hubrig, S., Sch\"{o}ller, M., Schnerr, R.S., 
                          Gonz\`{a}lez, J.F., Ignace, R., \& Henrichs, H.F.
                          2008, A\&A, 490, 793
                          
\bibitem[Hubrig et al.(2009a)]{Hubetal09a} Hubrig, S., Steltzer, B., Sch\"{o}ller, M. et al. 2009a, AN, 330, 708

\bibitem[Hubrig et al.(2009b)]{Hubetal09b} Hubrig, S., Briquet, M., De
  Cat, P. et al. 2009b, AN, 330, 317

\bibitem[Hubrig et al.(2009c)]{Hubetal09c} Hubrig, S., Mathys, G. ,
  Kurtz, D. W. et al. 2009c, MNRAS, 396, 1018

\bibitem[Kochukhov \& Bagnulo(2006)]{KocBag06} Kochukhov, O., Bagnulo,
S. 2006, A\&A 450, 763

%%%\bibitem[Kolenberg \& Bagnulo(2009)]{KolBag09} Kolenberg, K., \& Bagnulo, S. 2009, A\&A, 498, 543

\bibitem[K\"{u}nzli et al.(1997)]{Kunetal97} K\"{u}nzli, M., North, P., Kurucz, R. L., Nicolet, B. 1997, A\&AS, 122, 51

\bibitem[Kurtz(1982)]{Kur82} Kurtz, D. W. 1982, MNRAS, 200, 807

\bibitem[Izzo et al.(2010)]{Izzetal10} Izzo, C., de Bilbao, L., Larsen, J., et al. 
                                       2010, SPIE, 7737, 773729
%bibitem[Izzo et al.(2011)]{Izzetal11} Izzo, C., de Bilbao, L., Larsen, J.M. 2011
%                          FORS Pipeline User Manual, Issue 4.1, VLT-MAN-ESO\_19500-4106

%%%\bibitem[Jordan et al.(2005)]{Joretal05} Jordan, S., Werner, K., \& O'Toole, S.J. 
%%%                          2005, A\&A, 432, 273

\bibitem[Landstreet et al.(1975)]{Lanetal75} Landstreet, J. D., Borra,
E. F., Angel, J. R. P., Illing, R. M. E. 1975, ApJ, 201, 624

\bibitem[Landstreet(1982)]{Lan82} Landstreet, J. D. 1982, ApJ, 258, 639
                          
\bibitem[Landstreet \& Mathys(2000)]{LanMat00} Landstreet, J. D., Mathys, G. M. 2000, A\&A, 359, 213

\bibitem[Landstreet et al.(2007)]{Lanetal07} Landstreet, J. D., Bagnulo, S., Andretta, V. et al. 2007, A\&A, 470, 685

\bibitem[Landstreet et al.(2012)]{Lanetal12}Landstreet, J. D.,
Bagnulo, S., Valyavin, G. G. et al. 2012, A\&A 545, A30

\bibitem[Leroy et al.(1994)]{Leretal94} Leroy, J.-L., Bagnulo, S., Landolfi, M., Landi degl'Innocenti, E. 1994, A\&A, 284, 174

\bibitem[Mathys(1989)]{Mat89} Mathys, G. 1989, Fund. Cosm. Phys., 13, 143

\bibitem[Mathys(1991)]{Mat91} Mathys, G. 1991, A\&AS, 89, 121

\bibitem[Mathys \& Hubrig(1997)]{MatHub97} Mathys, G., Hubrig S. 1997, A\&AS, 124, 475

\bibitem[Mathys et al.(1997)]{Matetal97} Mathys, G., Hubrig, S.,
  Landstreet, J. D., Lanz, T., \& Manfroid, J. 1997, A\&AS, 123, 353

\bibitem[Mathys \& Hubrig(2006)]{MatHub06} Mathys, G., Hubrig,
  S. 2006, A\&A, 453, 699

%%%\bibitem[McSwain(2008)]{McS08} McSwain, M.V. 2008, ApJ, 686, 1269

\bibitem[Mermilliod et al.(1997)]{Meretal97} Mermilliod, J.-C., Mermilliod, M. Hauck, B. 1997, A\&AS, 124, 349

\bibitem[Napiwotzki et al.(1993)]{Napetal93} Napiwotzki, R., Sch\"{o}nberner, D., Wenske, V. 1993, A\&A, 268, 653

\bibitem[Netopil et al.(2008)]{Netetal08} Netopil, M., Paunzen, E.,
Maitzen, H. M., North, P., Hubrig, S. 2008, A\&A, 491, 545

%%%\bibitem[O'Toole et al.(2005)]{Otoetal05} O'Toole, S.J., Jordan, S., Friedrich, S., \& Heber, U.
%%%                          2005, A\&A, 437, 227

\bibitem[Preston \& Pyper(1965)]{PrePyp65} Preston, G. W., Pyper,
D. M. 1965, ApJ, 142, 983 

\bibitem[Shultz et al.(2012)]{Shuetal12} Shultz, M., Wade, G. A.,
Grunhut, J. et al. 2012, ApJ, 750, 2

\bibitem[Shulyak et al.(2010)]{Shuetal10} Shulyak, D., Ryabchikova,
T., Kildiyarova, R., Kochukhov, O. 2010, MNRAS, 520A, 88S

\bibitem[Shulyak et al.(2013)]{Shuetal13} Shulyak, D., Ryabchikova,
T., Kochukhov, O. 2013, A\&A, 551, A14

%%%\bibitem[Silvester et al.(2009)]{Siletal09} Silvester, J., Neiner, C., Henrichs, H.F., et al.
%%%                          2009 MNRAS, 398, 1505

\bibitem[Thompson(1983)]{Tho83} Thompson, I. B. 1983, MNRAS, 205, 43P

%%%\bibitem[Valyavin et al.(2008)]{Valetal08} Valyavin, G., Wade, G.A., Bagnulo, S., Szeifert, T.,
%%%                          Landstreet, J.D., Han, Inwoo, \& Burenkov, A.
%%%                          2008, ApJ, 683, 466

\bibitem[van den Heuvel(1971)]{vandenHeu71} van den Heuvel,
E. P. J. 1971, A\&A, 11, 461

\bibitem[Wade et al.(2000)]{Wadetal00} Wade, G. A., Donati, J.-F., Landstreet, J. D., \& Shorlin, S. L. S. 2000, MNRAS, 313, 851

%%%\bibitem[Wade et al.(2005)]{Wadetal05}	Wade, G.A., Drouin, D., Bagnulo, S., et al.
%%%                          2005, A\&A, 442, 31L

%%%\bibitem[Wade et al.(2007)]{Wadetal07} Wade, G.A., Bagnulo, S., Drouin, D., Landstreet, J.D., \& Monin, D.
%%%                          2007, MNRAS, 376, 1145
                          
\bibitem[Wolff \& Bonsack(1972)]{WolBon72} Wolff, S. C., Bonsack,
W. K. 1972, ApJ, 176, 425

\bibitem[Wolff(1975)]{Wol75} Wolff, S. C. 1975, ApJ, 202, 127

\bibitem[Wolff \& Hagen(1976)]{WolHag76} Wolff, S. C., \& Hagen, G. 1976, PASP, 88, 119

\end{thebibliography}
\end{document}